\documentclass[showpacs,preprintnumbers,prb,amsmath,amssymb,superscriptaddress,twocolumn]{revtex4}
\usepackage{endnotes}
\usepackage{dcolumn}
\usepackage{bm}
\usepackage{graphicx}
\usepackage{color}
\usepackage{amssymb}

\begin{document}

\preprint{APS/123-QED}

\title{The magnetic ground state properties of non-centrosymmetric CePt$_3$B$_{1-x}$Si$_x$} 

\author{D. Rauch}
\author{P. Horenburg}
\author{S. Hartwig}
\author{F. J. Litterst}
\author{S. S\"ullow}
\affiliation{Institute of Condensed Matter Physics, Technische Universit\"at Braunschweig, D-38106 Braunschweig, Germany}

\author{H. Luetkens}
\author{C. Baines}
\affiliation{Laboratory for Muon Spin Spectroscopy, Paul Scherrer Institute, CH-5232 Villigen, Switzerland}

\author{S. Yamazaki}
\author{H. Hidaka}
\author{H. Amitsuka}
\affiliation{Graduate School of Science, Hokkaido University, Sapporo 060-0810, Japan}

\author{E. Bauer}
\affiliation{Institute of Solid State Physics, Vienna University of Technology, A-1090 Vienna, Austria}
\date{\today}

\begin{abstract}
We present a study of the alloying series of the non-centrosymmetric $f$--electron intermetallic CePt$_3$B$_{1-x}$Si$_x$ by means of muon spin rotation and relaxation measurements. In addition, we include a high pressure magnetization investigation of the stoichiometric parent compound CePt$_3$B. From our data we establish the nature of the magnetic ground state properties of the series, derive the  ordered magnetic moment as function of stoichiometry and gain insight into the evolution of the symmetry of the ordered magnetic state with $x$. We thus can verify the notion that the behavior of the sample series can essentially be understood within the framework of the Doniach phase diagram. Further, our findings raise the issue of the role the Dzyaloshinskii-Moriya magnetic interaction plays in correlated electron materials, and its effect on magnetic fluctuations in such materials.
\end{abstract}


\maketitle
\noindent

Throughout the past three decades in the field of studies on heavy fermion superconductors topics of prime interest are the mechanisms of the pairing and the symmetry of the superconducting state \cite{Sigrist1991,Pfleiderer2009}. Especially for the spin triplet state in unconventional superconductors there are strong arguments that the binding of electrons to Cooper pairs occurs via spin fluctuations \cite{Mathur1998}. In this context, CePt$_3$Si is the first heavy fermion superconductor with a lack of inversion symmetry of its crystallographic lattice \cite{Bauer2004}. This causes a spin-orbit splitting of the Fermi surface, which might generate chiral spin states \cite{Hashimoto2004}. A pure spin triplet pairing in such non-centrosymmetric system is excluded because of fundamental symmetry arguments \cite{Anderson1984}. Now, CePt$_3$Si exhibits an anomalously large superconducting upper critical field B$_{c2} \approx 5$\,T \cite{Bauer2004}. To resolve this conflict, nowadays the common view is that the superconducting pairing involves an admixture of a spin-singlet and triplet state. Still, up to now, a comprehensive explanation of superconductivity in these non-centrosymmetric systems is lacking.

Therefore, CePt$_3$Si has been the focus of very intensive research efforts in recent years \cite{Bauer2004,Takeuchi2007,Nicklas2010,Aoki2010}. An unconventional heavy fermion ($\gamma = 0.39$ J/mole K$^2$) superconducting ground state has been established below T$_c=$ 0.75\,K (0.45\,K in a high quality single crystal \cite{Takeuchi2007}; the discrepancy has not been resolved conclusively as yet). This superconducting state is in coexistence with a long-range antiferromagnetically ordered state, with an ordering wave vector ${\bf q} = (0, 0, 0.5)$ of strongly reduced magnetic moments $\mu_{ord} = 0.16\,\mu_B$/Ce being detected below the N$\mathrm{\acute{e}}$el temperature $T_N = 2.2$\,K \cite{Metoki2004,Amato2005,Kaneko2012}. 

In contrast, the isostructural system CePt$_3$B does neither show superconductivity nor heavy fermion behavior at low temperatures \cite{Suellow1994}. Both, CePt$_3$Si and CePt$_3$B, crystallize in the tetragonal non-centrosymmetric CePt$_3$B structure (space group $P4mm$) at ambient pressure with lattice parameters $a = 4.072/4.003$\,\AA~ and $c = 5.442/5.075$\,\AA~ for the Si/B compound, respec\-ti\-vely \cite{Bauer2004,Suellow1994,Lackner2005}. In a previous paper, we have argued that the combination of lattice parameter difference (corresponding to chemical pressure) and electron count difference (Si contributes one more electron than B to the conduction band) can be viewed to lead to an {\it effective chemical pressure} \cite{Rauch2012A}. In result, the experimentally observed physical properties of CePt$_3$B lead to the conclusion that this material is a local moment magnet with much weaker electronic correlations than CePt$_3$Si, and thus would represent a low pressure variant of CePt$_3$Si. 

Based on thermodynamic and transport experiments it has been established that CePt$_3$B undergoes two magnetic transitions at low temperatures, the first one into a state of essentially antiferromagnetic (AFM) nature below $T_N = 7.8$\,K, the latter one into a state with a weakly ferromagnetic (FM) signature below $T_C \sim 4.5 - 6$\,K \cite{Suellow1994,Lackner2005}. To account for the two magnetic phases in CePt$_3$B one line of reasoning would be that there is a transition of large magnetic moments (order of magnitude $\sim \mu_B$) into an antiferromagnetic structure below $T_N$, which transforms into a weakly ferromagnetic one below $T_C$ through canting of the magnetic moments. Within this line of thought, the canting could be a consequence of the lacking inversion symmetry, as this gives rise to an additional magnetic exchange term, the Dzyaloshinskii-Moriya (DM) interaction \cite{Dzyaloshinsky1958,Moriya1960}. All in all, in CePt$_3$B, a combination of ferro-, antiferromagnetic or orthogonal couplings in the lattice might thus produce complex magnetic states such as for instance canted or helical ones \cite{Bak1980,Nakanishi1980,Prokes2011}.

Surprisingly, in a study of the magnetically ordered phases of CePt$_3$B by means of neutron scattering and $\mu SR$ this scenario could not be verified \cite{Rauch2012B}. On the one hand, in $\mu SR$ experiments both transitions at $T_N$ and $T_C$ have been identified as bulk transitions. As well, the muon precession frequency does suggest the presence of a large ordered magnetic moment ($\sim 1 \mu_B$/Ce) in both phases. On the other hand, in neutron powder diffraction no additional intensity from scattering in the magnetically ordered phase has been observed. As yet, this failure to detect magnetic intensity in neutron scattering is not understood. 

By studying the series CePt$_3$B$_{1-x}$Si$_x$, $0 \leq x \leq 1$, via thermodynamic measurements and electronic transport we have previously established that there is a relationship between the magnetically ordered phases in both compounds. In result, the increasing electron correlations with replacement of B by Si lead to the disappearance of the weakly ferromagnetic phase of CePt$_3$B, while the antiferromagnetic one persists. In this situation, the aim of the present study is to characterize the magnetic phase diagram in CePt$_3$B$_{1-x}$Si$_x$ in further detail by microscopic techniques and high pressure experiments, in order to explore the correlation between different types of magnetic order with superconductivity in CePt$_3$Si.

The samples of the series CePt$_3$B$_{1-x}$Si$_x$, $0 \leq x \leq 0.8$, have been characterized extensively regarding their structural and physical properties in the Refs. \cite{Rauch2012A,Rauch2012B}. Here, we extend these studies by means of $\mu$SR experiments performed on CePt$_3$B$_{1-x}$Si$_x$ in weak transverse applied field and zero field using the general purpose surface-muon spectrometer (GPS) of the Swiss Muon Source at the Paul Scherrer Institute, Villigen. Additionally, $\mu$SR experiments on CePt$_3$B$_{0.2}$Si$_{0.8}$ were carried out with the low temperature facility (LTF) instrument allowing to reach temperatures between $0.019 - 1.6$\,K. The samples CePt$_3$B$_{1-x}$Si$_x$ are available in polycrystalline form ($x=0.2$ and $0.4$) or as powder ($x=0.6$ and $0.8$).

To supplement our findings we have in addition performed high pressure magnetization experiments on CePt$_3$B. This way, we can directly verify the notion set out in Ref. \cite{Rauch2012A} of CePt$_3$B being a low pressure variant of CePt$_3$Si. Then, it will be possible to put the studies on CePt$_3$B$_{1-x}$Si$_x$ into a broader context, relating the Doniach phase diagram and the Dzyaloshinskii-Moriya interaction with the strength of electronic correlations.

Two types of pressure experiments were carried out, first using a CuBe clamp cell for pressures up to about 1\,GPa, secondly a miniature zirconia anvil cell for pressures up to about 5\,GPa \cite{Tateiwa2011}. The experiments using the CuBe clamp follow the experimental procedure set out in Ref. \cite{Kreitlow2005}, those for the anvil cell in Ref. \cite{Tateiwa2014}.

\begin{figure}[t!]
	\centering
		\includegraphics[width=1\linewidth]{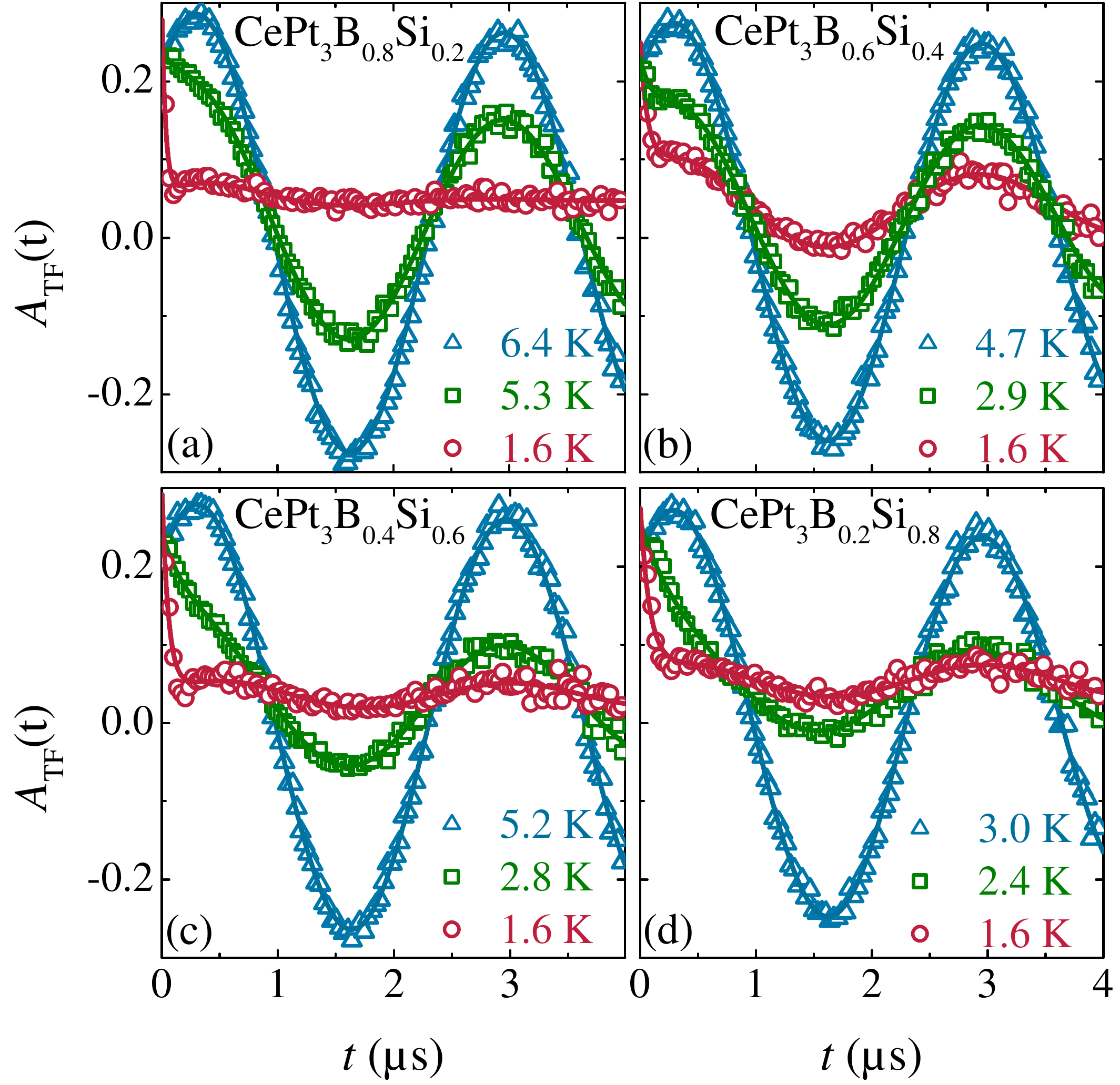}
	\caption{Weak transverse field $\mu$SR asymmetry spectra at various temperatures in an external field of $3 $\,mT for CePt$_3$B$_{1-x}$Si$_x$, $x= 0.2$ (a), $0.4$ (b), $0.6$ (c) and $0.8$ (d). Solid lines are fits to the data, for details see text.}
	\label{fig:TFmusr}
\end{figure}

First, we present weak transverse field (wTF) $\mu$SR experiments CePt$_3$B$_{1-x}$Si$_x$, $0.2\leq x \leq 0.8$, in the temperature range of $1.6 - 300$\,K using an external field of $3$\,mT applied at an angle of $90^\circ$ relative to the polarized muon spin. Representative wTF asymmetry spectra are depicted in Fig.~\ref{fig:TFmusr}. To illustrate the typical behavior, selected measurements in the paramagnetic regime (blue), close to the phase transition (green) and in the magnetically ordered state (red) are displayed. Qualitatively, all samples show a similar behavior. In the paramagnetic regime ($T>$ $T_N$), the transverse external magnetic field leads to a weakly damped muon precession with frequency $\nu \sim B_{ext}$. At the magnetic phase transition ($T \approx T_N$) the muons start to see an additional internal field from the spontaneous magnetic ordering ($B_{int} \gg B_{ext}$). This results in the observation of an initial damping of the muon oscillation in the asymmetry signal. Due to the polycrystalline nature of the materials studied, the internal field is oriented differently for each crystal grain. Consequently, the resulting fields at the muon sites are inhomogeneous which lead to an incoherent precession. In the long-range antiferromagnetically and weakly ferromagnetically ordered state ($T \ll T_N$) the muon precession from the external field is almost completely suppressed. The asymmetry spectra now acquire the character of an exponential decay function.

The muon precession $A_{\mathrm{TF}}(t)$ can be described by a summation taking into account the signal in the paramagnetic phase (first term) and a magnetically ordered contribution (second term):
\begin{equation}
	A_{\mathrm{TF}} (t) = a_1\cos \left( \omega t + \phi \right) e^{\lambda_1 t} +a_2e^{-\lambda_2 t}.
	\label{eq:TFCe}
\end{equation}
The harmonic oscillation resulting from the external field in the paramagnetic state is given by a frequency $\nu=\omega/2\pi\sim 0.37$\,MHz and a phase $\phi \sim 41^\circ$. The magnetic ordering manifests itself by an exponential decay with the decay parameter $\lambda_2$. Finally, after full suppression of the oscillation in the paramagnetic phase a signal offset of about $a_{res}/a_{tot} \sim 0.05-0.2$ can be identified in the magnetically ordered state, reflecting a signal contribution from the experimental device. The total asymmetry is determined to $a_{tot} = a_1 + a_2 + a_{res}$. Altogether, the time dependent asymmetry spectra from the wTF experiments are fitted with Eq.~\ref{eq:TFCe} and included as solid lines in Fig.~\ref{fig:TFmusr}.

\begin{figure}[t!]
	\centering
		\includegraphics[width=1\linewidth]{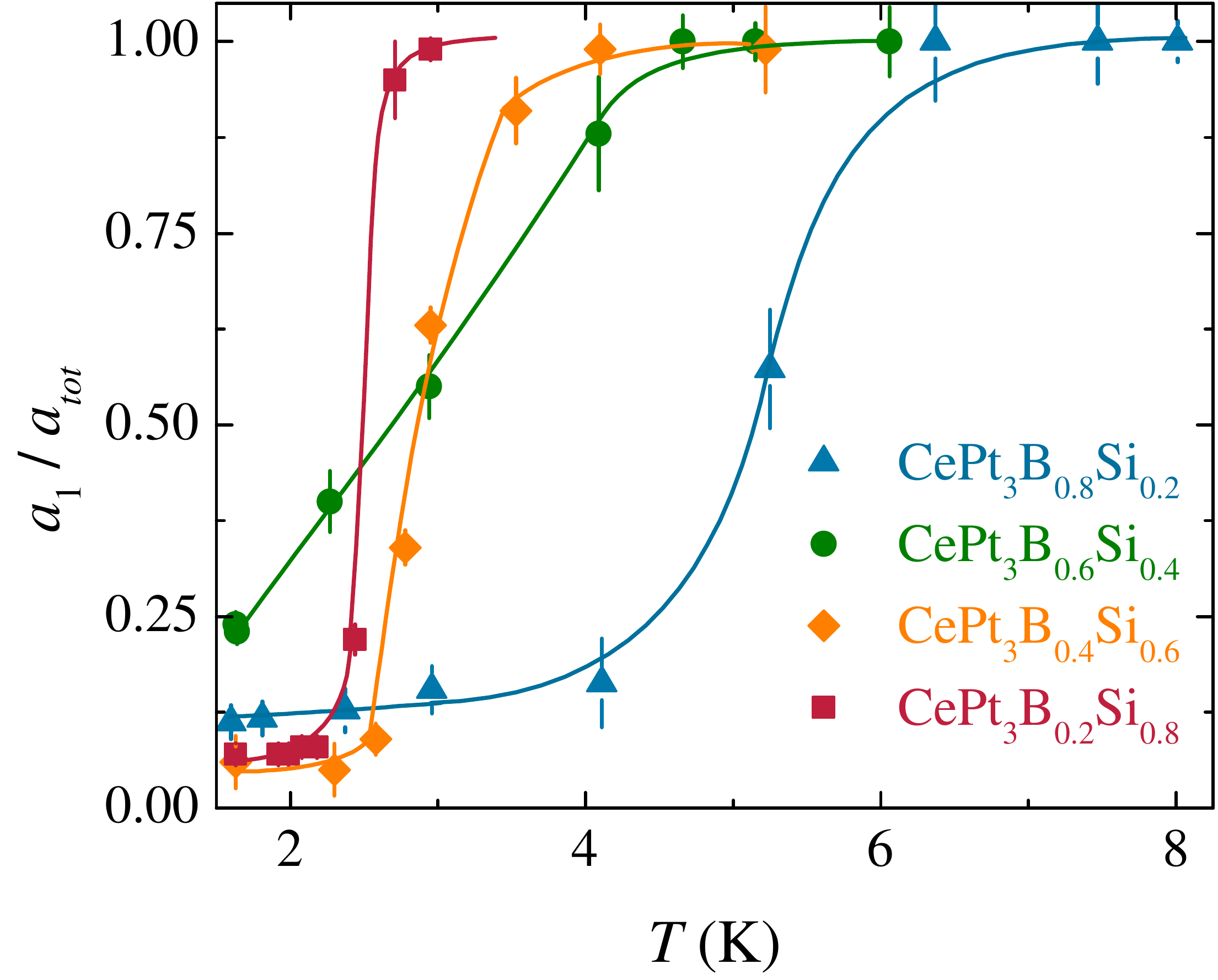}
	\caption{Temperature dependency of the normalized asymmetry parameter $a_1/a_{tot}$ from wTF measurements in CePt$_3$B$_{1-x}$Si$_x$, $0.2 \leq x \leq 0.8$, lines are guides to the eye.}
	\label{fig:asymmetry}
\end{figure}

The transition from the paramagnetic regime, where $a_1/a_{tot}\cong1$, into the magnetically ordered state is reflected in a temperature dependence of the asymmetry parameters, with $a_2$ increasing on behalf of $a_1$ upon transition into the magnetically ordered state. The amplitude of the oscillating signal component represented by $a_1$ is proportional to the paramagnetic volume fraction. Hence, $a_1$ for each sample serves as an indicator for the volume fraction not magnetically ordered. Correspondingly, Fig.~\ref{fig:asymmetry} depicts the temperature dependence of the normalized asymmetry parameter $a_1/a_{tot}$, as obtained from fits of Eq.~\ref{eq:TFCe} to the data. From this figure, the transition into the antiferromagnetic state can be determined from the onset of the decrease of the asymmetry to $T_N=6.1$\,K in CePt$_3$B$_{0.8}$Si$_{0.2}$, $4.4$\,K in CePt$_3$B$_{0.6}$Si$_{0.4}$, $3.8$\,K in CePt$_3$B$_{0.4}$Si$_{0.6}$ and $2.7$\,K in CePt$_3$B$_{0.2}$Si$_{0.8}$ within an experimental error of $\sim0.5$\,K, in good agreement with the results from the bulk studies described in Ref.~\cite{Rauch2012A}.

The samples CePt$_3$B$_{1-x}$Si$_x$, $x =$ 0.2, 0.6 and 0.8 exhibit a fairly sharp transition with a decrease from $a_1/a_{tot}\cong1$ in the paramagnetic regime to $a_1 \leq 0.1\,a_{tot}$ at the lowest temperatures. This proves the bulk nature of the magnetically ordered phases for these samples, with a volume fraction of the magnetically ordered phase larger than $90$\,\% for all samples. In contrast, the sample CePt$_3$B$_{0.6}$Si$_{0.4}$ shows a broadened transition and a magnetic volume fraction of $\sim80\,$\% at lowest temperatures, which is in line with the residual weak oscillatory asymmetry signal observed at very low temperatures (Fig.~\ref{fig:TFmusr}b). Also, these observations correspond to the temperature dependent magnetic specific heat contribution $C_{mag}/T$, which also reveals a broadened transition into the antiferromagnetic state for this sample \cite{Rauch2012A}.

More detailed information on the magnetic properties, and in particular the nature of the magnetically ordered phases, are obtained from zero magnetic field (ZF) $\mu$SR experiments on CePt$_3$B$_{1-x}$Si$_x$, $0.0 \leq x \leq 0.8$. In this experimental configuration, the muon behavior only reflects the muon precession and relaxation from the internal magnetic fields in the magnetically ordered phases. Then, features like magnetic phase transitions or spin reorientation processes will result in changes of the internal magnetic field and show up in the muon signal.

\begin{figure}[t!]
	\centering
		\includegraphics[width=1\linewidth]{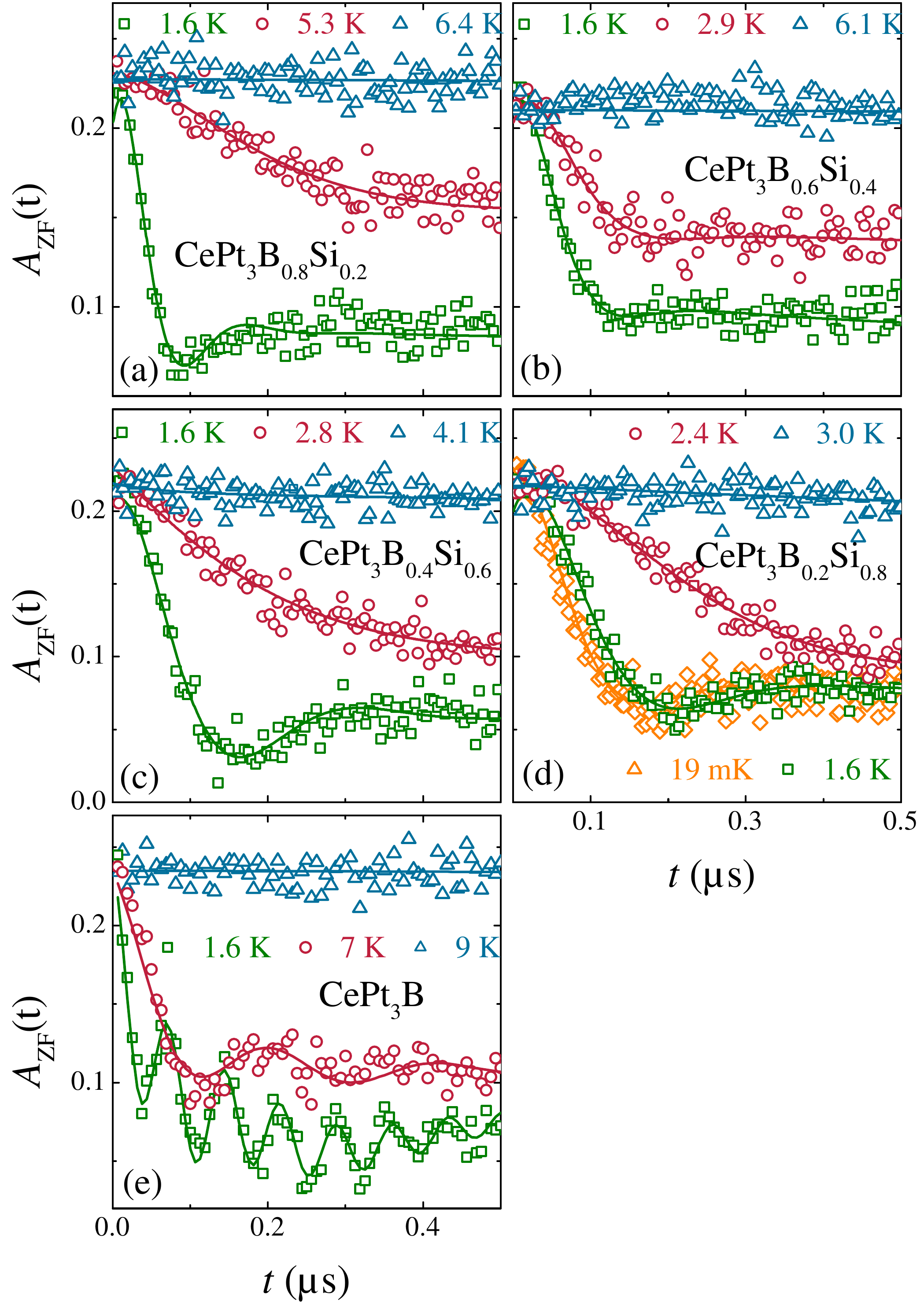}
	\caption{Zero field $\mu$SR asymmetry spectra at various temperatures for CePt$_3$B$_{1-x}$Si$_x$, $x = 0.2$ (a), $0.4$ (b), $0.6$ (c), $0.8$ (d) and $0.0$ from Ref.~\cite{Rauch2012B} (e). Solid lines are fits to the data. For $x = 0.8$ normalized LTF values at $T = 19$\,mK are included.}
	\label{fig:ZFmusr}
	\end{figure}

In Fig.~\ref{fig:ZFmusr} time dependent ZF $\mu$SR asymmetry spectra for CePt$_3$B$_{1-x}$Si$_x$, $0.0 \leq x\leq 0.8$ for selected temperatures are summarized. For all samples, the transition from the paramagnetic phase into the magnetically ordered state is signaled by the occurrence of a spontaneous damped muon oscillation, which for $x > 0.0$ fully decays over one oscillation period. In agreement with the wTF $\mu$SR measurements, CePt$_3$B$_{0.6}$Si$_{0.4}$ is the only sample with only a weakly discernible oscillation (Fig.~\ref{fig:ZFmusr}b). Assuming that this sample contains a secondary non-magnetic phase with a volume fraction of about 20\,\%, this behavior can be explained by a superposition of the oscillatory muon signal from the main phase and a background signal from a non-magnetic minority phase. 

Furthermore, to investigate the possible presence of a ferromagnetic contribution in CePt$_3$B$_{0.2}$Si$_{0.8}$ additional ZF $\mu$SR experiments are carried out using the LTF instrument for temperatures between $0.019 - 1.6$\,K, with one data set shown as an example in Fig.~\ref{fig:ZFmusr}d. In order to compare the $\mu$SR data obtained from the GPS and LTF facilities, $A_{\mathrm{ZF}} (t)$ of LTF experiments are normalized (and adjusted to the values from the GPS) to correct for the signal from the silver backing used in the LTF measurements.

Following previous studies on CePt$_3$Si and CePt$_3$B \cite{Amato2005,Rauch2012B}, for CePt$_3$B$_{1-x}$Si$_x$, $0.0 \leq x \leq 1.0$, the muon signal in the magnetically ordered phases is described as a superposition of damped oscillation signals. Correspondingly, the ZF $\mu$SR asymmetry signal in the magnetically ordered state is analyzed by a superposition of $n$ terms, were $n$ represents the number of distinct muon sites assumed for the fit:
\begin{eqnarray}
	A_{\mathrm{ZF}} (t) = \sum_i^n a_i \left\{\alpha_i \cos(\omega_i t+\phi_i) \exp\left(-\lambda_{T,i} t\right)\right.\nonumber \\ \left.+ (1-\alpha_i)\exp\left(-\lambda_{L,i} t\right)\right\}.
	\label{eq:ZFCe}
\end{eqnarray}
Here, the first component describes the muon precession with a frequency $\nu_i=\omega_i/2\pi$ caused by the local internal magnetic field $B_{int}$ at each muon site. The coefficient $\alpha_i$ denotes the fraction of transverse internal field components of the field distribution with respect to the initial muon spin, which give rise to a precession, and is expected to be 2/3 for a random magnetic environment. For the temperature dependent data analyzed here a nearly constant fraction $\alpha_i \approx 0.7 \pm 0.1$ was found, indicating an almost coherent orientation for the spatial average of the localized Ce moments. As expected for a polycrystalline sample, the second "1/3 term" of each summand represents the fraction of muons possessing an initial polarization along the the internal field direction. Further, the longitudinal depolarization rate $\lambda_{L,i}$ ($\sim 0.1\,\mu s^{-1}$) reflects solely internal spin dynamics, while the transverse depolarization rate $\lambda_{T,i}$ ($\sim 10 - 20\,\mu s^{-1}$) describes both static and dynamic effects like the spin-spin-interaction. The temperature dependence of the precession frequency $\omega_i$ reflects the evolution of the (sublattice) magnetization.

For higher temperatures $T > T_N$, the paramagnetic phase is best described by the asymmetry function
\begin{equation}
	A_{\mathrm{ZF,PM}} (t) = a_{KT} G_{KT}^{dyn}(\nu,\sigma,\Gamma,t),
	\label{eq:ZFCePM}
\end{equation}
with the dynamic Kubo-Toyabe function $G_{KT}^{dyn}(\nu,\sigma,\Gamma,t)$, the field distribution $\sigma$, the hopping rate $\Gamma$ and the frequency $\nu$ (Refs.~\cite{Hayano1979,Dalmas1992,Keren1994}). In ZF $\mu$SR, the frequency is fixed to $\nu=0$. The parameter $\Gamma$ describes dynamic effects, which are associated with hopping processes of muons between different interstitial sites. For a single muon site, the prefactors $a_{i}$ and $a_{KT}$ in the Eqs.~\ref{eq:ZFCe} and \ref{eq:ZFCePM} represent the asymmetry parameters, with the sum of these parameters giving the total asymmetry $a_{tot}$. 

\begin{figure}[t!]
	\centering
		\includegraphics[width=1\linewidth]{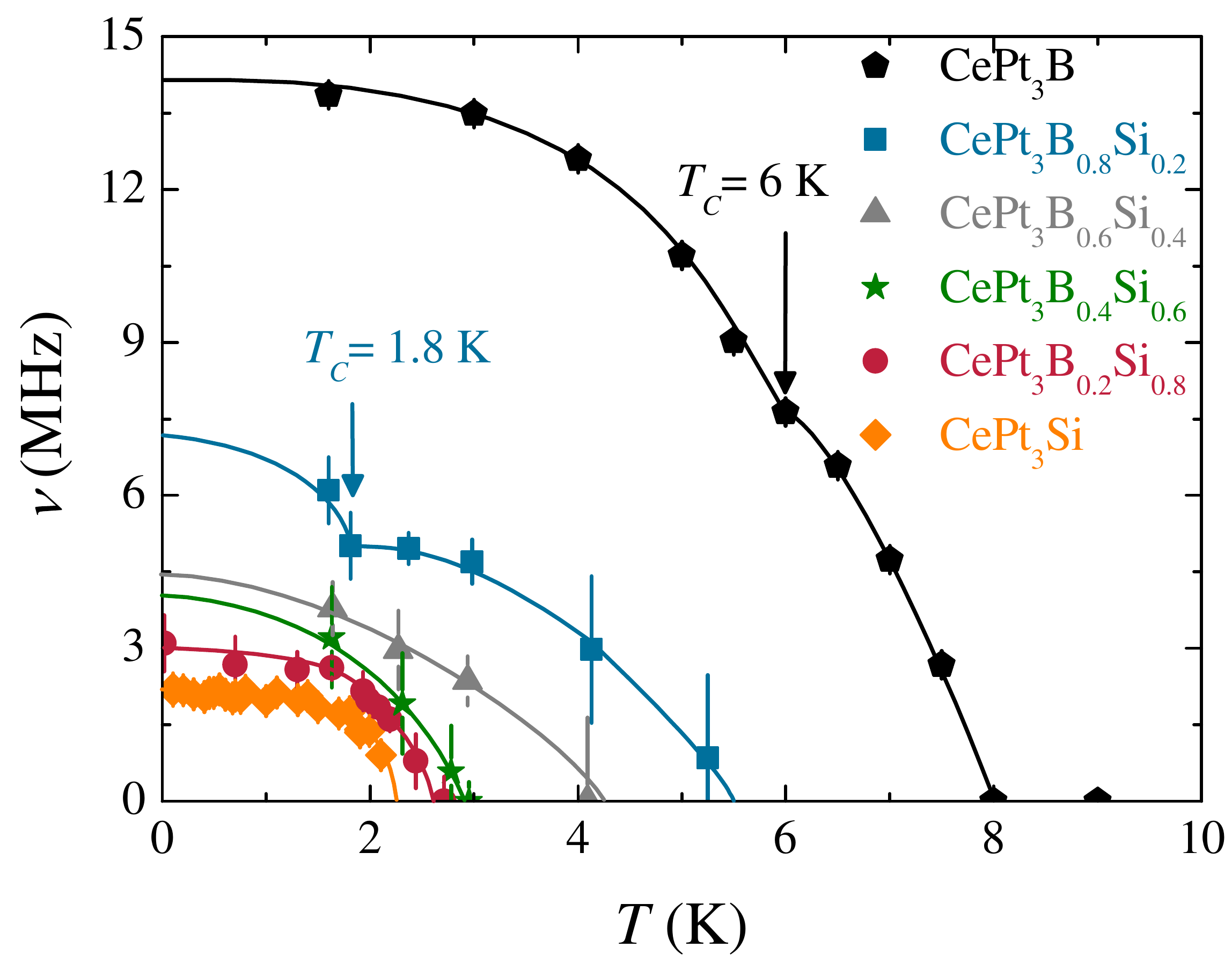}
	\caption{Temperature dependence of ZF $\mu$SR frequencies of CePt$_3$B$_{1-x}$Si$_x$, $0.0 \leq x \leq 1.0$. The data for $x = 0.0$ are taken from Ref.~\cite{Rauch2012B}, for $x = 1.0$ from Ref.~\cite{Amato2005}. Lines are included for better visibility of the transition temperatures $T_N$ and $T_C$, arrows mark $T_C$.}
	\label{fig:Cefrequency}
\end{figure}

Previously, the experimental ZF muon depolarization data for CePt$_3$B has been analyzed using Eq.~\ref{eq:ZFCe} with a superposition of three terms ($n$ = 3) for the magnetically ordered state and Eq.~\ref{eq:ZFCePM} for the paramagnetic phase \cite{Rauch2012B}, implying the existence of three distinct muon sites in the material. In these fits, the phase $\phi = 0$ and $\lambda_L \sim 0.1\,\mu s^{-1}$ are fixed for all muon sites. This procedure allows to study in detail the temperature dependence of the various fit parameters. Far below the phase transition, for each muon site the transverse damping rate $\lambda_{T,i}$ is almost constant and diverges at the phase transitions. The temperature dependence of the precession frequencies $\omega_i$ reflect the evolution of the bulk/sublattice magnetization in the antiferromagnetic and weakly ferromagnetic states and is depicted in Fig.~\ref{fig:Cefrequency} for the largest frequency. 

Also, the antiferromagnetically ordered state in CePt$_3$Si has been analyzed using this approach, but with a sum of two muon precession signals indicating the presence of at least two inequivalent muon stopping sites sensing very low internal magnetic fields ($\leq 2.3\,$MHz), cf.~Ref.~\cite{Amato2005} (data from that reference are included in Fig.~\ref{fig:Cefrequency}). 

All experimental ZF muon depolarization data for CePt$_3$B$_{1-x}$Si$_x$, $0.2 \leq x \leq 0.8$, could be analyzed using Eq.~\ref{eq:ZFCe} with a superposition of two terms ($n = 2$) for the ordered state and Eq.~\ref{eq:ZFCePM} for the paramagnetic phase, and which are depicted as solid lines in Fig.~\ref{fig:ZFmusr}. In comparison, the necessity to use three muon sites for CePt$_3$B reflects that the damping of the muon asymmetry spectra is much weaker in CePt$_3$B than for silicon concentrations of $x \geq 0.2$. In effect, about ten oscillation periods have to be taken into account in the fit of the CePt$_3$B data, which can only be properly done by using three muon sites (see Fig.~\ref{fig:ZFmusr}). Per individual muon site, however, the magnetic fields, and thus frequencies, are comparable. Conversely, for the alloyed samples the strong damping effectively wipes out the information about different local fields, {\it viz.}, muon sites, and the magnetic field associated to the precession frequency $\nu$ in Eq.~\ref{eq:ZFCe} represents an average field. Now, the muon precession frequencies $\nu = \gamma_ \mu / 2 \pi B_{int}$ are of particular interest, which reflect the evolution of the magnetically ordered phases and magnetic moment $\mu_{ord}$ as function of temperature and silicon concentration. Fig.~\ref{fig:Cefrequency} displays the temperature dependent highest precession frequencies $\omega_i$ for CePt$_3$B$_{1-x}$Si$_x$, $0.0 \leq x \leq 1.0$. 

From the temperature dependence of the muon precession frequencies, the N$\mathrm{\acute{e}}$el temperature $T_N$ can be estimated by determining $T(\nu \rightarrow 0)$. The maximum internal fields $B_{int}$ are derived from the values $\nu(T=0)$ as well. Both quantities are summarized in Tab.~\ref{tab:Temp4}.

\begin{table}[b!]
	\centering
	  \begin{tabular}{l c c c }\hline  
		 ~$x$~  &  ~$T_N$\,(K)~ & ~$T_C$\,(K)~ & ~$B_{int}(T=0$\,K)\,(mT)~ \\
		\hline
		\hline   
		$0.0$ & $8.0$ & $6.0$ & $104$\\
		$0.2$ & $5.5$ & $1.8$ & $53$\\
		$0.4$ & $4.3$ & $< 1.6$ & $32$\\
		$0.6$ & $3.0$ & $< 1.6$ & $32$\\
		$0.8$ & $2.7$ & $< 0.019$ & $23$\\
		$1.0$ & $2.3$ & - & $16$\\
    \hline
		\end{tabular}
\caption{Magnetic transition temperatures $T_N$ and $T_C$ and maximal internal fields $B_{int}(T=0$\,K) of CePt$_3$B$_{1-x}$Si$_x$, $0.0\leq x\leq 1.0$, determined from ZF $\mu$SR measurements. The values for CePt$_3$Si are taken from the Ref.~\cite{Amato2005}.}
	\label{tab:Temp4}
\end{table}

Overall, we observe that the $\mu$SR frequencies decrease with increasing Si substitution. This behavior is in agreement with the observations on the magnetic bulk properties in Ref.~\cite{Rauch2012A}. Further, for decreasing temperature, we found an initially rapid increase of precession frequencies, which for $T \rightarrow 0$ becomes almost temperature independent. Equally, $T_N$ and $B_{int}(T=0$~K) decrease with increasing Si concentration. For $0.0 \leq x \leq 0.4$ both quantities decrease significantly with Si alloying, while for larger $x$ values the dependency is much weaker and for $x \rightarrow 1.0$ it approaches the behavior of CePt$_3$Si \cite{Amato2005}.

CePt$_3$B$_{0.6}$Si$_{0.4}$ exhibits a transition into the antiferromagnetic phase at $4.3$~K, whereas the specific heat $C_{mag}(T)$ only displays a weak signature at this temperature \cite{Rauch2012A}. Again, this might be attributed to a lower quality of this particular sample, consistent with wTF $\mu$SR and bulk experiments. Summarizing the findings so far, the suppression of the antiferromagnetic phase in CePt$_3$B$_{1-x}$Si$_x$ with increasing silicon amount fully confirms the bulk measurements.

In contrast to the detection of long-range antiferromagnetic ordering, which is reflected in the occurrence of a distinct spontaneous muon precession, determining the transition into the weakly ferromagnetic phase is a much harder task. For CePt$_3$B it has previously been demonstrated that a change of slope in the temperature dependent muon oscillation frequency occurs below $T_C$ \cite{Rauch2012B}. Ferromagnetic ordering causes additional muon oscillation compared to the antiferromagnetic muon precession (shown as green line in Fig.~\ref{fig:ZFmusr}e). 

\begin{figure}[t!]
	\centering
		\includegraphics[width=1\linewidth]{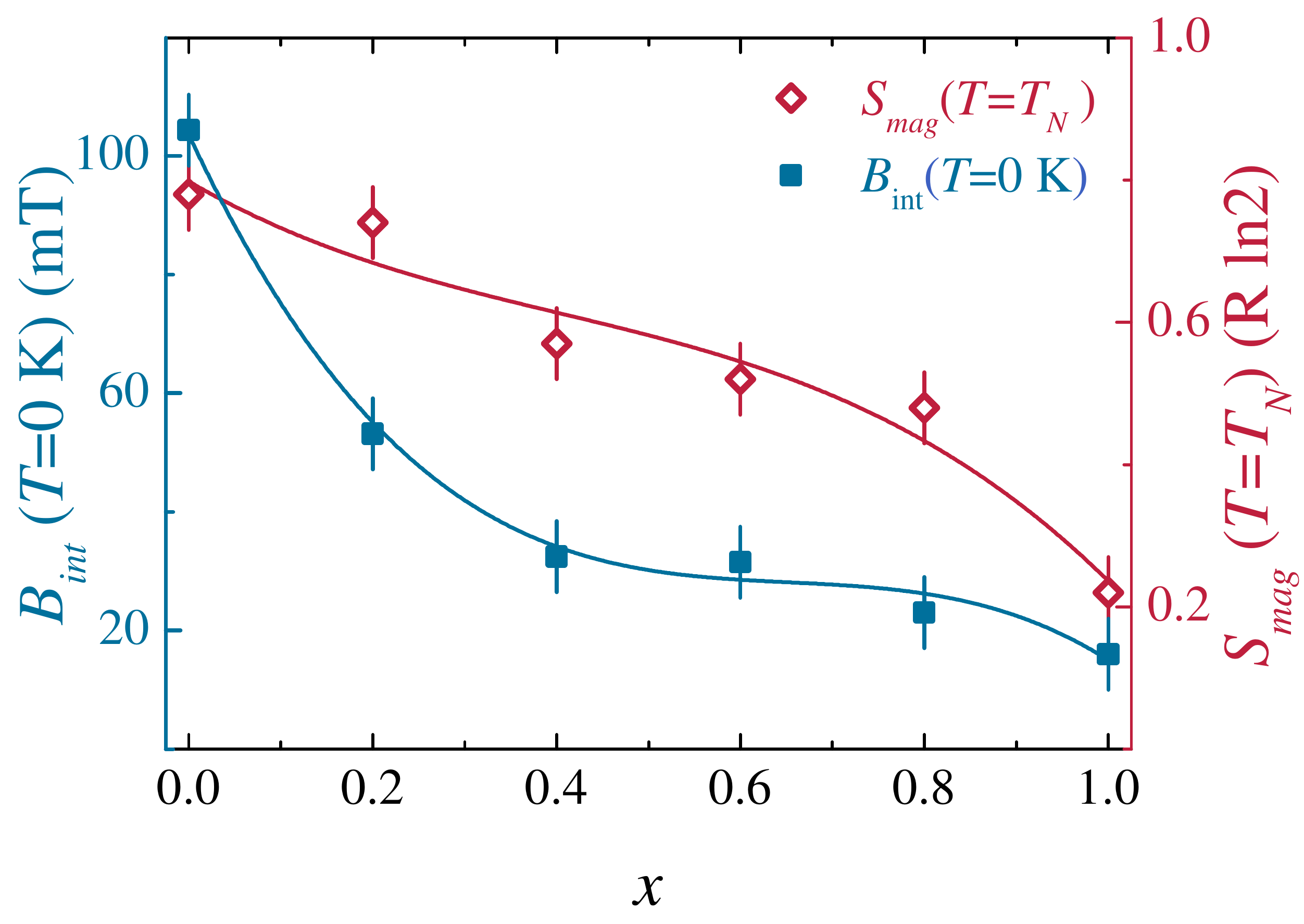}
	\caption{Silicon concentration dependency of the maximum internal magnetic field $B_{int}(T=0$\,K) and magnetic entropy $S_{mag}(T=T_N)$ for CePt$_3$B$_{1-x}$Si$_x$, $0.0\leq x\leq 1.0$. Solid lines are guides to the eye.}
	\label{fig:fieldvsentropie}
\end{figure}

Further inspection of the temperature dependent oscillation frequencies reveals that also for CePt$_3$B$_{0.8}$Si$_{0.2}$ there is a feature similar to that seen in CePt$_3$B. This behavior can be interpreted as an evidence for the occurrence of weak ferromagnetic ordering from canted magnetic moments, which we identified as $T_C$ (see arrows in Fig.~\ref{fig:Cefrequency}, values included in Tab.~\ref{tab:Temp4}). The Curie temperature $T_C$ obtained from these kinks are in agreement with the bulk measurements (Tab.~\ref{tab:Temp4}). CePt$_3$B$_{1-x}$Si$_x$ with $x = 0.4$ and 0.6 do not show a clear indication for ferromagnetic behavior for $T \geq 1.6$\,K in the GPS experiments. Hence, from our data we set an upper limit for the ferromagnetic transition temperatures for these samples of $T_C \leq 1.6$\,K. 

CePt$_3$B$_{0.2}$Si$_{0.8}$ is also investigated in the LTF facility. The depolarization of the LTF measurements does not show a clear-cut feature signaling ferromagnetic behavior down to temperatures of 19\,mK. As well, in the experiment no indication for superconductivity is observed. 

The size of the magnetically ordered moment $\mu_{ord}$ represents one central characteristic of the magnetic properties of CePt$_3$B$_{1-x}$Si$_x$, which can be estimated from the internal field $B_{int}(T=0$\,K) $\propto \mu_{ord}$. A qualitative comparison with the magnetic entropy $S_{mag}(T=T_N)$ from Ref.~\cite{Rauch2012A} reveals that both behave in a similar way (Fig.~\ref{fig:fieldvsentropie}). The magnetic entropy is a measure of disorder of the magnetic lattice, which is proportional to $\mu_{ord}$. Increasing the silicon concentration leads to a strictly monotone decrease of $B_{int}(T=0$\,K) and $S_{mag}(T=T_N)$, reflecting the gradual suppression of magnetic ordering with Si. Qualitatively, this can be understood within the Doniach phase diagram as a result of the increasing screening of localized magnetic moments via the Kondo effect.

Semiquantitatively, Fig.~\ref{fig:fieldvsentropie} further indicates that the size of the magnetically ordered moment changes by about a factor of about 5 from CePt$_3$B to CePt$_3$Si. Given the experimentally determined moment of CePt$_3$Si, $\mu_{ord} = 0.16\, \mu_B$/Ce atom \cite{Metoki2004}, for CePt$_3$B we would expect a moment of the order of $\mu_{ord} \sim 0.8\, \mu_B$/Ce atom, fully consistent with the bulk data. 

\begin{figure}[t!]
\centering
	\includegraphics[width=1\linewidth]{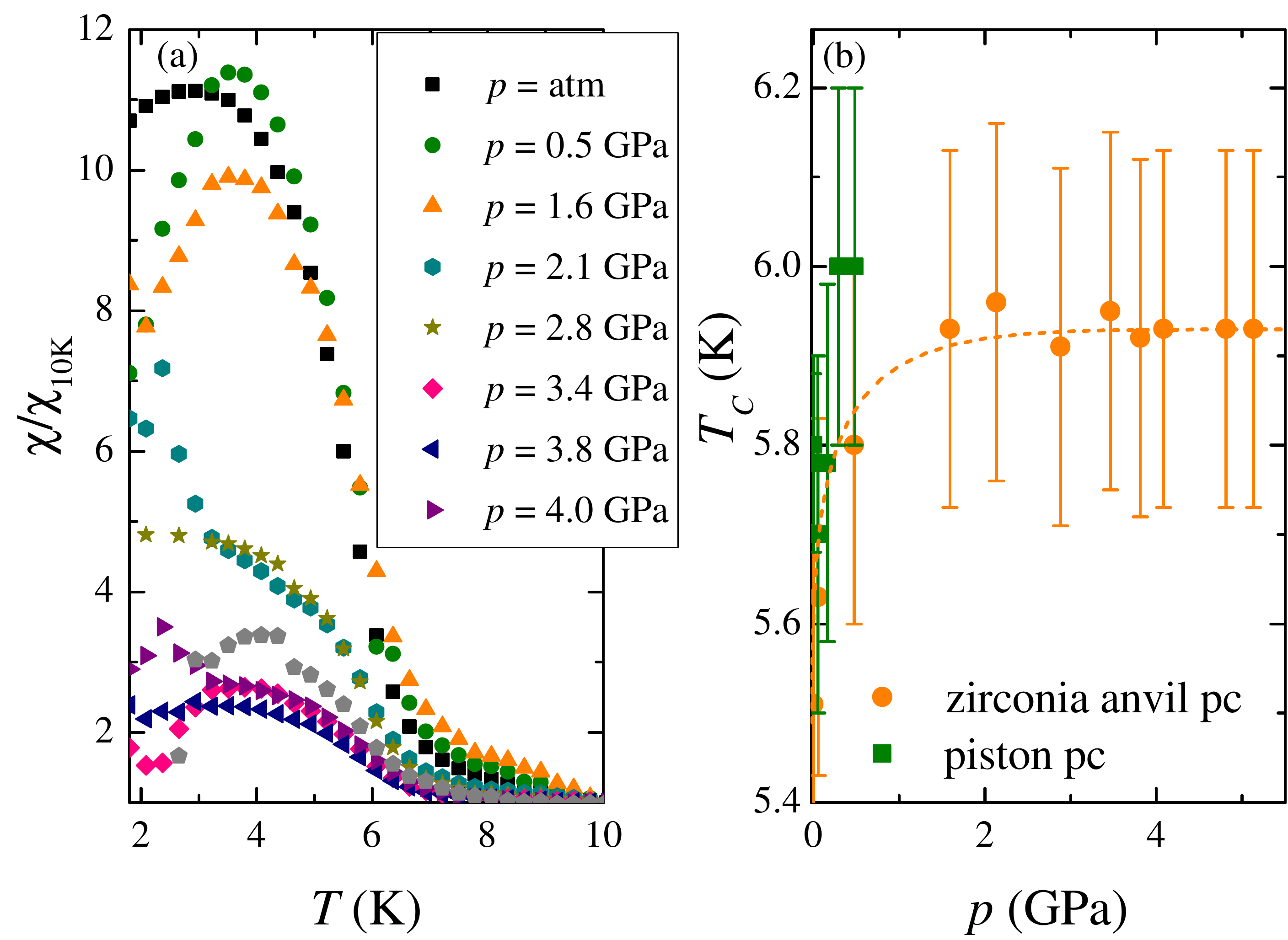}
	\caption{(a) Temperature dependence up to 10\,K of the susceptibility $\chi/\chi_{10K}$ for various pressure values from the zirconia anvil pressure cell experiment on CePt$_3$B. (b) Pressure dependency up to $p = 5.5$\,GPa of the ferromagnetic transition temperature $T_C$ of CePt$_3$B investigated at low temperatures in a piston pressure cell (green squares) and a zirconia anvil pressure (orange circles) cell. The dashed line construction is a guide to the eye. }
	\label{fig:pressure}
\end{figure} 

As set out above, the antiferromagnetic state of CePt$_3$B undergoes a transition producing a weakly ferromagnetic behavior below $T_C=5.6$\,K. In addition to above microscopic studies, the accompanying typical ferromagnetic signature in the susceptibility $\chi$ is investigated by studying its pressure dependence up to $5.5$\,GPa (Fig.~\ref{fig:pressure}a). As in CePt$_3$Si~\cite{Motoyama2008} it is observed that overall $\chi$ of CePt$_3$B decreases with pressure within the experimental error. The antiferromagnetic transition of CePt$_3$B can not be seen clearly in this experiment. 

From the experimental data, the pressure dependency of the ferromagnetic transition temperature $T_C$ is determined as the maximum of the derivative of $\chi T$. This procedure gives a rather large experimental error as result of the small sample weight as compared to the large background. In numbers, the zirconia anvil pressure cell gives rise to a magnetic background signal of about $\mu \sim 10^{-4}$ emu, which constitutes $\sim 95$\,\% of the signal including a sample with a mass of $m$(CePt$_3$B)$= 0.22$\,mg. The overall evolution of the pressure dependence of $T_C$ determined from both the piston pressure cell (green circles) and zirconia anvil pressure cell experiment (orange squares) is depicted in Fig.~\ref{fig:pressure}b. Up to $0.55$\,GPa, the transition temperature of the weakly ferromagnetic phase first slightly increases with pressure, and subsequently saturates at highest pressures. Such behavior is in accordance with that expected from the Doniach phase diagram in the regime of well defined local magnetic moments \cite{Doniach1977} ({\it viz.}, for relatively weak hybridization).  

\begin{figure}[t!]
	\centering
		\includegraphics[width=1\linewidth]{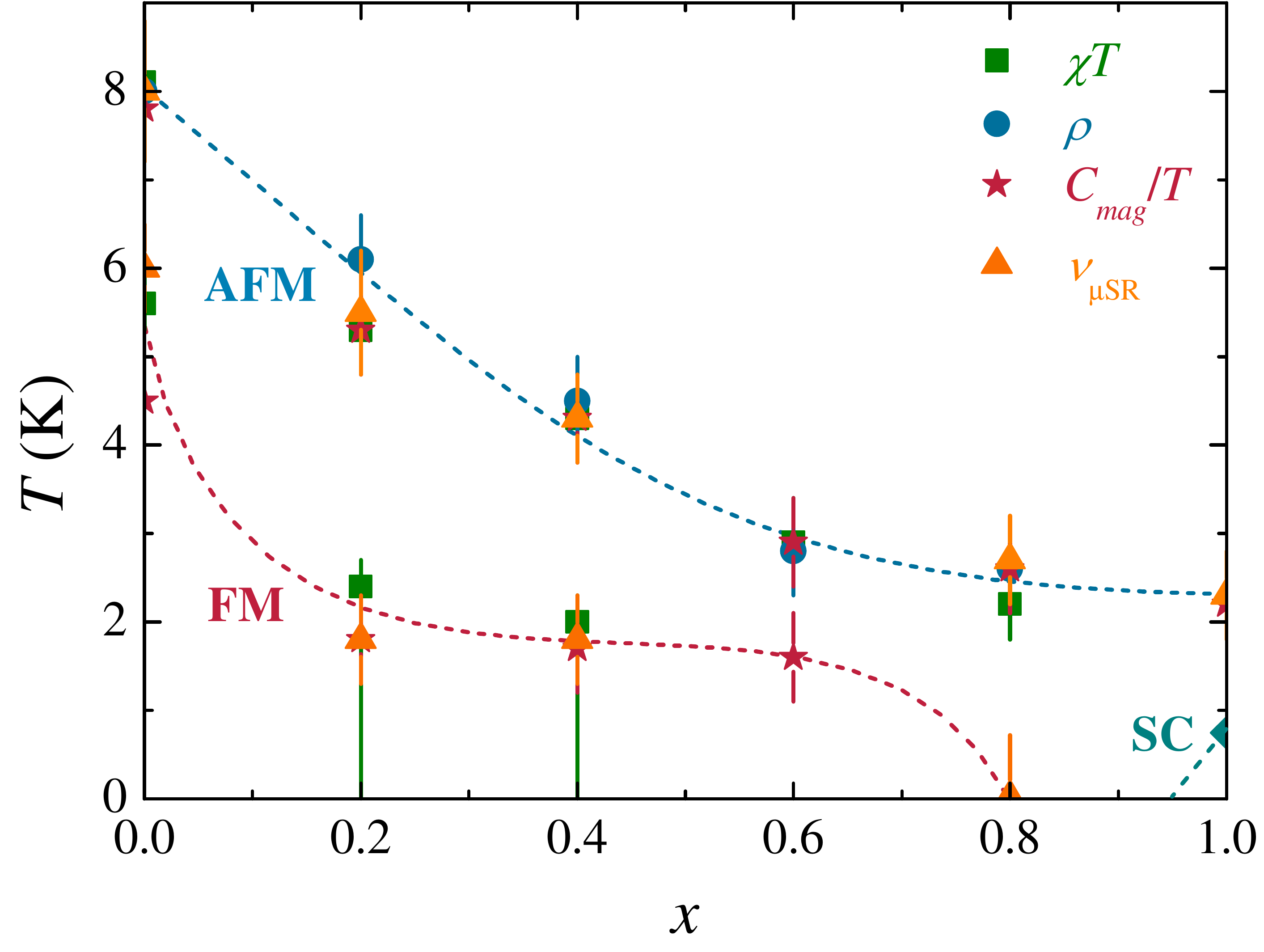}
	\caption{Magnetic phase diagram of the antiferromagnetic (AFM) $T_N$, ferromagnetic (FM) $T_C$ and superconducting (SC) $T_c$ transition temperatures as function of silicon concentration $x$ for CePt$_3$B$_{1-x}$Si$_x$, $0.0 \leq x \leq 1.0$; data for $x = 0.0$ and $1.0$ are taken from Refs.~\cite{Suellow1994,Bauer2004}.}
	\label{fig:phasedia}
\end{figure}

Our findings are fully consistent with the results of a resistivity study under pressure up to 1.85\,GPa by Lackner {\it et al.} \cite{Lackner2005}, who observe an increase of the antiferromagnetic transition $T_N$ by about 1\,K in this pressure range. Altogether, these observations reflect that the weakly ferromagnetic phase, and correspondingly the antiferromagnetic one, are still in the local moment region of the Doniach phase diagram. Conversely, a much higher pressure would be required to drive the system into a range of strong electronic correlations and close to a magnetic instability.

To summarize the findings presented here on the alloying system CePt$_3$B$_{1-x}$Si$_x$, $0.0 \leq x \leq 1.0$, antiferromagnetic and weakly ferromagnetic phases are observed by means of bulk pressure techniques and microscopic $\mu$SR experiments. The effect of silicon alloying is demonstrated in the corresponding magnetic phase diagram (Fig.~\ref{fig:phasedia}), where the transition temperatures obtained from bulk and microscopic experimental techniques are plotted as function of $x$. While there is some variation of the absolute values of the antiferromagnetic $T_N$ and ferromagnetic transition temperature $T_C$, overall the different experiments yield rather similar results regarding the phase diagram. 

All in all, the phase diagram implies that the antiferromagnetic phase in CePt$_3$B ($T_N=7.8$~K) transforms continuously into that in CePt$_3$Si ($T_N=2.2$~K), with a smooth suppression of the ordering temperatures. The antiferromagnetic transition temperatures $T_N$ are almost constant for a silicon content of $x \geq 0.6$. In contrast, the weakly ferromagnetic phase in CePt$_3$B ($T_C=5.6$~K) is completely suppressed at a critical value of $x_c \approx 0.8$. Superconductivity appears close to stoichiometric CePt$_3$Si ($T_c=0.75$~K). 

Weakly ferromagnetic behavior can occur as a result of the DM interaction in non-centrosymmetric systems by canting of antiferromagnetically ordered spins. This ferromagnetic signature is almost completely suppressed at the critical concentration $x_c$, suggesting that the DM interaction is weakened with Si alloying. With the complete suppression of the DM interaction, unconventional superconductivity occurs in CePt$_3$Si. Qualitatively, this behavior can be discussed within the concept of the Doniach phase diagram, which considers the competition of long-range magnetic order from an RKKY-like exchange and the Kondo effect \cite{Doniach1977}. The suppression of antiferromagnetic and weakly ferromagnetic order are accompanied by a significant enhancement of electronic correlations, as evidenced by the increasing electronic specific heat coefficient $\gamma$. Here, it could be argued that in the strongly correlated state the DM interaction becomes less relevant, as it is not well-defined for a complex delocalized and correlated state of $f$--electrons coupled to the conduction electron bath.

\begin{figure}[t!]
	\centering
		\includegraphics[width=1\linewidth]{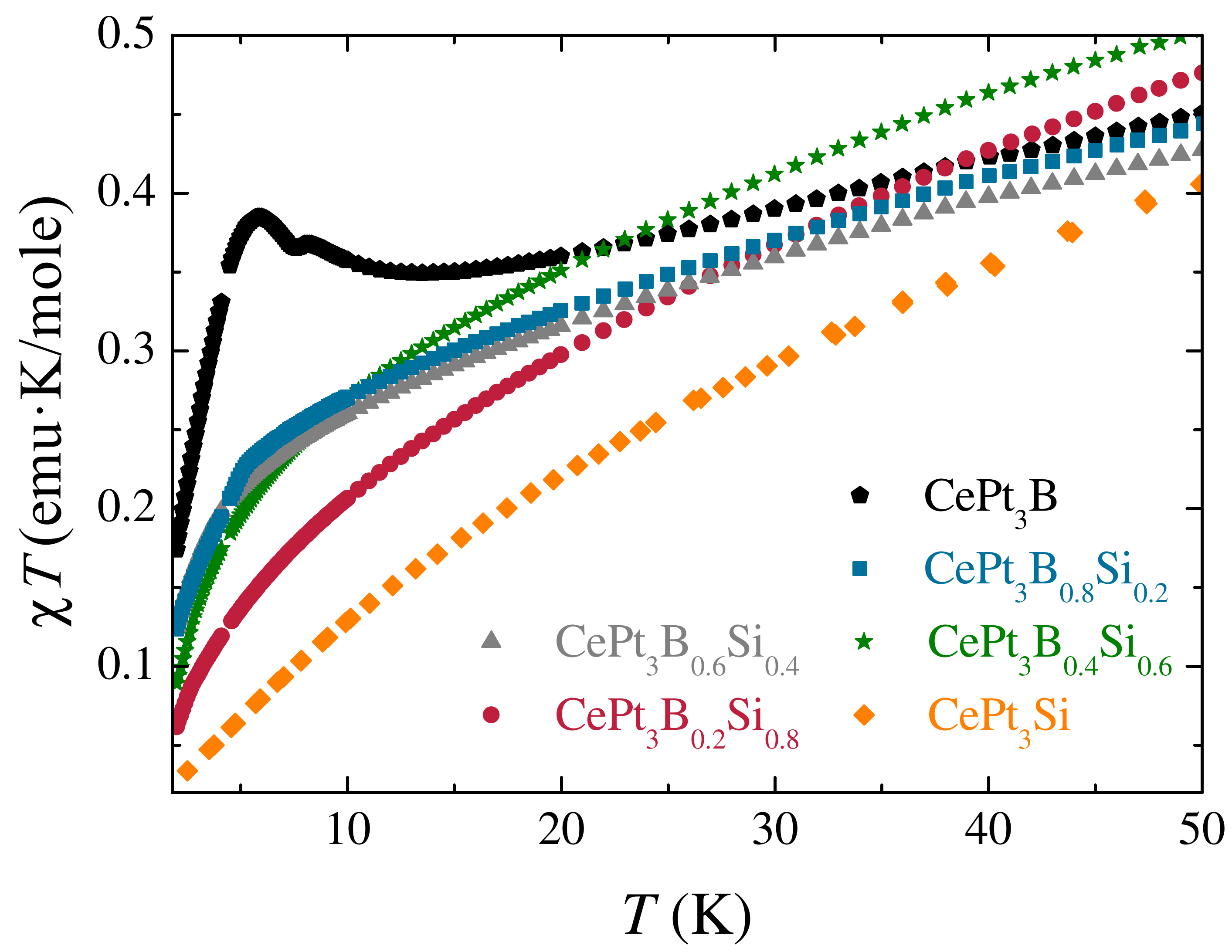}
	\caption{A plot of the product $\chi T$ versus $T$ to illustrate the nature of predominant magnetic fluctuations for CePt$_3$B$_{1-x}$Si$_x$, $0 \leq x \leq 1$; data for $x = 1$ have been taken from Ref. \cite{Bauer2005}.}
	\label{fig:chit}
\end{figure}

With these ideas about the relevance of the DM interaction, we can speculate about the transformation of the antiferromagnetic phase in CePt$_3$B into that of CePt$_3$Si. As indicated in the introduction, the antiferromagnetic structure in CePt$_3$Si is a rather simple one, with a wave vector ${\bf q} = (0, 0, 0.5)$ \cite{Metoki2004}. While for CePt$_3$B the wave vector is unknown so far, neutron scattering experiments have verified that it is not the one of CePt$_3$Si \cite{Rauch2012B}. Further, in these experiments no scattering intensity from a magnetically ordered phase has been seen in a scattering angle range that would correspond to ''ordinary'' wave vectors, for instance with a doubling of the unit cell etc.. Conversely, this seems to suggest that the ordered magnetic phase in CePt$_3$B might be one with a long-range modulation of the spin-periodicity, which might lead to a signal in neutron scattering at low scattering angles inaccessible in the experiment carried out in Ref.~\cite{Rauch2012B}. Such a phase would also be consistent with the presence of the DM interaction in CePt$_3$B, which frequently produces such long-range modulations in magnetic materials. With the suppression of the DM interaction, the long-range modulated state disappears, and instead the system locks into the wave vector of CePt$_3$Si. 

In this context, a common approach to qualitatively assess the relevance and nature of magnetic fluctuations is to plot the product of susceptibility and temperature, $\chi T$, against the temperature $T$. This procedure is carried out in Fig. \ref{fig:chit} for the complete alloying series CePt$_3$B$_{1-x}$Si$_x$. A decreasing product $\chi T$ with decreasing $T$ is considered to indicate predominant antiferromagnetic fluctuations, while conversely $\chi T$ decreasing with increasing $T$ signals a ferromagnetic character of the fluctuations. From the figure, it is evident that CePt$_3$B shows a signature of predominant ferromagnetic fluctuations, which is consistent with the phase diagram. Upon alloying, the figure suggests that these ferromagnetic fluctuations are suppressed on behalf of antiferromagnetic ones. With this observation in mind, a line of thought might be that this change of the character of the fluctuation spectrum is an element relevant to account for superconductivity in CePt$_3$Si.

Regarding the starting point of our study, considering CePt$_3$B as a low pressure variant of CePt$_3$Si, we have performed pressure experiments up to $5.5$\,GPa on CePt$_3$B which show a slight increase of $T_C$ by about 5\,\% with a constant ferromagnetic contribution at a pressure $1$\,GPa\,$ \geq p \geq 5$\,GPa. In contrast to CePt$_3$Si, in which the magnetic phase is suppressed at $p = 0.6$\,GPa, in CePt$_3$B a pressure of $5.5$\,GPa is not sufficient to destroy the ferromagnetic behavior. Within the concept of the Doniach model, this observation reflects that CePt$_3$B is still deep in the local moment region of the Doniach phase diagram. Thus, a much higher pressure would be required to drive the system into the range of strong electronic correlations and close to a magnetic instability. Consequently, it would be very interesting to see if the properties of the alloyed samples CePt$_3$B$_{1-x}$Si$_x$ under very high pressure resemble those of CePt$_3$Si, and in particular if these samples become superconducting.

Part of this work was supported by the Japanese Society for the Promotion of Science.

\end{document}